# Neutrino oscillations and Lorentz invariance violation

**S. A. Alavi, M. Dehghani Madise**
Department of physics, Hakim Sabzevari university, P, O, Box 397, Sabzevar, Iran

Lorentz invariance is a well known fundamental concept of special relativity but its violation is predicted by some variations of quantum gravity, string theory, and some alternatives to general relativity. We study neutrino oscillation at the presence of the Lorentz invariance violation in vacuum, normal and nuclear media. It is shown that for very high energy neutrinos the neutrino oscillations due to mass difference is suppressed and its contribution to the conversion probability is negligibly small, but the contribution of the Lorentz invariance violation term is not zero. This means that the observation of very high energy neutrino oscillations may be a signal of the discreteness of space and Lorentz invariance violation.

## 1. Introduction

Neutrino was first postulated in 1930 by Wolfgang Pauli to preserve the conservation of energy, momentum, and angular momentum in beta decay. Neutrino oscillation was predicted first by Bruno Pontecorvo in 1957. It is a quantum mechanical phenomenon and implies that neutrino created with a specific lepton flavor (electron, muon, or tau) can later be changed to different flavor. Neutrino oscillation is of great theoretical and experimental interest, particularly, the existence of neutrino oscillation resolved the long-standing solar neutrino problem. Neutrino oscillation was discovered by Super-Kamiokande and the Sudbury Neutrino Observatories.

On the other hand, Heisenberg's uncertainty principle is one of the most important results of twentieth century physics which states that, there is a fundamental limit for the measurement accuracy, with which certain pairs of physical observables, such as position and momentum or energy and time, can not be measured, simultaneously. Recentlly, deduced from different approaches related to the quantum gravity (QG) such as black hole physics and string theory, a minimal length was predicted. Such a consequence arises due to the fact that strings cannot probe distances smaller than the string scale. This view is also supposed by many Gedanken experiments. The existence of a minimal length leads to the modification of the Heisenberg uncertainty principle to a Generalized Uncertainty Principle (GUP), which has various implications on a wide range of physical systems.

Lorentz invariance is one of the main and basic concepts in special relativity (for more than one century) and states that, the laws of physics are invariant under Lorentz transformation. Lorentz invariance imply that spacetime structure is the same at all scales, that is, Lorentz symmetry assumes a scale-free nature of spacetime. there is no fundamental length scale associated with the Lorentz group. So the discreteness of space can lead to the violation of the Lorentz invariance (VLI). As mentioned above, the existence of fundamental length is predicted by some theories including, quantum gravity, string theory, Doubly Special Relativity and noncommutative geometry. An interesting point is that a generalized uncertainty principle (GUP) leads to the quantization of length, areas and volumes [1,2]. These results suggest the discreteness of space and fundamental granular structure of space.

So there is possibility of Lorentz invariance violation (LIV) in nature due to discreteness of space. Then a question naturally arise : Can we check Lorentz invariance violation by neutrino oscillations? . In this work we study neutrino oscillation in a discrete space and compare the results with ordinary (continuous space ) to check deviation from Lorentz invariance.

### 2.1. Generalized Uncertainty Principle (GUP)

The GUP relation which is consistent with string theory and black hole physics has the following form [1,2] :

$$[x_i, p_j] = i\hbar[\delta_{ij} - \alpha\left(p\delta_{ij} + \frac{p_i p_j}{p}\right) + \alpha^2(p^2 \delta_{ij} + 3 p_i p_j)] \quad (1)$$

It is shown in [1,2], that this GUP commutation relation implies the discreteness of space. For a free particle, Dirac equation is :

$$i\frac{\partial \psi}{\partial t} = (\beta mc^2 + c\vec{\alpha}.\vec{p})\psi \tag{2}$$

The GUP-corrected energies of the Dirac equation [2] are given by :

$$E^2 = (\hbar k)^2 + (mc^2)^2 \tag{3}$$

where :

$$k = k_0 + \alpha \hbar k_0^2 \tag{4}$$

and $k_0$ is the wavenumber that satisfies in the relation :

$$E_0^2 = (\hbar k_0)^2 + (mc^2)^2 \tag{5}$$

which is the free Dirac particle energy in the absence of GUP.
For high energy neutrinos we have :

$$E = k_0 + \frac{(mc^2)^2}{2k_0} + 2\alpha\hbar k_0 \tag{6}$$

## 2.2. Neutrino oscillations- Basic equations.

The probability of measuring a particular flavor for a neutrino varies periodically as it propagates through space. Flavor neutrinos are weak interaction eigenstates and can be considered as the superposition of the mass eigenstates. The basic equations governing the neutrino oscillations are as follows :

$$H|v_i> = E_i|v_i> \tag{7}$$

i=1,2,3. $|v_i>$ are the mass eigenstates. The flavor eigenstates and the mass eigenstates are related by a unitary transformation as follows :

$$|v_l> = \sum_{i=1}^{3} U_{li} |v_i> \tag{8}$$

where $l$ is the flavor index. Time evolution of the flavor neutrino states are as follows :

$$|v_l(t)> = e^{-iHt}|v_l> = \sum_{i=1}^{2,3} U_{li} e^{-iE_i t} |v_i> \tag{9}$$

The probability of oscillations is given by :

$$P_{v_l \to v_{l'}} = |\langle v_{l'} || v_l \rangle|^2 \tag{10}$$

For two component neutrino oscillations i=1,2. The unitari matrix U is given by :

$$\begin{pmatrix} \cos\theta & \sin\theta \\ -\sin\theta & \cos\theta \end{pmatrix} \tag{11}$$

So, we have :

$$\begin{aligned}|v_e> &= \cos\theta |v_1> + \sin\theta |v_2> \\ |v_\mu> &= -\sin\theta |v_1> + \cos\theta |v_2>\end{aligned} \tag{12}$$

## 2.3. Neutrino oscillations due to LIV in a discrete space: vacuum

The Hamiltonian in the mass eigenstates is :

$$H_{vaccum} = \begin{pmatrix} E_1 & 0 \\ 0 & E_2 \end{pmatrix} \approx E + 2\hbar \begin{pmatrix} \alpha_1 E_1 & 0 \\ 0 & \alpha_2 E_2 \end{pmatrix} + \begin{pmatrix} \frac{m_1^2}{2E} & 0 \\ 0 & \frac{m_2^2}{2E} \end{pmatrix} + \frac{1}{16} \begin{pmatrix} B_1^2 a^2 E^3 & 0 \\ 0 & B_2^2 a^2 E^3 \end{pmatrix} \quad (13)$$

where $\alpha_i$ are the GUP parameters in Eq.(1) and $B_i$ are free parameters characterizing the deviation from Lorentz symmetry [4-6]. The Hamiltonian in the base of flavor eigenstates is given by :

$$H = U H_{vaccum} U^\dagger = E + \frac{m_1^2 + m_2^2}{4E} + E^3 a^2 \left(\frac{B_1^2 + B_2^2}{16}\right) + \left[\left(\frac{\Delta m_{12}^2}{4E}\right) + E^3 a^2 \frac{\Delta B_{12}^2}{16} + 2\hbar \Delta(\alpha E)\right] \begin{pmatrix} -\cos 2\theta & \sin 2\theta \\ \sin 2\theta & \cos 2\theta \end{pmatrix} \quad (14)$$

where: $\Delta B_{12}^2 \equiv \Delta B^2 = B_2^2 - B_1^2$, and $\Delta(\alpha E) = \alpha_2 E_2 - \alpha_1 E_1$. The mixing angle is given by :

$$\tan 2\theta = \frac{2H_{12}}{H_{22} - H_{11}}$$

We consider the probability of oscillations (conversion) as follows [7]:

$$P(t) = (\sin 2\theta)^2 \sin^2(\varphi/2) \quad (15)$$

The phase $\varphi$ is given by :

$$\varphi = \int_{t_i}^{t} \epsilon(\tau) d\tau \quad (16)$$

where :
$$\epsilon = \frac{\Delta m_{12}^2}{2E} + \frac{1}{8} \Delta B_{12}^2 E^3 a^2 + 4\hbar \Delta(\alpha E) \quad (17)$$

So the oscllation phase can be separated into two parts :
$$\varphi = \varphi_{vac} + \varphi_{LIV} \quad (18)$$

Where $\varphi_{vac}$ is proportional to the first term $\frac{\Delta m_{12}^2}{2E}$ and $\varphi_{LV}$ is proportional to the rest terms. It is observed that for very high energy neutrinos (for which $\Delta m_{12}^2 \to 0$) the vacuum oscillation phase tends to zero and probability of oscillation is mainly contributed from LIV phase.

## 2.4. Neutrino oscillations due to LIV in a discrete space : Normal media

In normal media, due to the interaction of neutrinos with electrons, the Hamiltonian, the neutrino mixing angles and the mass squared are modified. Electron neutrinos can interact with electrons via both charge and neutral currents but muon and tau neutrinos interact with electrons only through charge current. In this case the Hamiltonian is given by :

$$H_{Normal} = E + \frac{m_1^2 + m_2^2}{4E} + \frac{\Delta m^2}{4E} \begin{pmatrix} -\cos 2\theta & \sin 2\theta \\ \sin 2\theta & \cos 2\theta \end{pmatrix} + 2\hbar \begin{pmatrix} \alpha_1 E_1 & 0 \\ 0 & \alpha_2 E_2 \end{pmatrix}$$
$$+ \begin{pmatrix} \sqrt{2} G_F n_e & 0 \\ 0 & 0 \end{pmatrix} - \frac{1}{\sqrt{2}} G_F n_n \quad (19)$$

See also [4]. $n_e$, $n_n$ and $G_F$ are the number densities of electrons, nucleons and the Fermi coupling constant respectively.

the energy ε is as follows :

$$\epsilon = \frac{\Delta m_{12}^2}{2E} - 2\sqrt{2} G_F n_e E + \frac{1}{8} \Delta B_{12}^2 E^3 a^2 + 4\hbar \Delta(\alpha E) \quad (20)$$

**eConf C16-09-04.3**

From Eqs. (16) and (20), we have the following form for the oscillation phase :

$$\varphi = \varphi_{vac} + \varphi_{mat} + \varphi_{LIV} \tag{21}$$

The osillation probability is given by :

$$P(t) = (\sin 2\theta)^2 \sin^2(\varphi/2) \tag{22}$$

It is observed that similar to the vacuum, for very high energy neutrinos the contribution of the first term in Eq. (20), vanishes but contrary to the vacuum there is a non-vanishing contribution due to the interaction of neutrinos with matter. There is also a contribution from discreteness of space and VLI term.

## 2.4. Neutrino oscillations due to LIV in a discrete space : Nuclear media

In nuclear media we need to consider the weak interaction between neutrinos and quarks . The Hamiltonian is as follows :

$$H_{Nuclear} = E + \frac{m_1^2 + m_2^2}{4E} + \frac{\Delta m^2}{4E}\begin{pmatrix} -\cos 2\theta & \sin 2\theta \\ \sin 2\theta & \cos 2\theta \end{pmatrix} + 2\hbar \begin{pmatrix} \alpha_1 E_1 & 0 \\ 0 & \alpha_2 E_2 \end{pmatrix}$$
$$+ \begin{pmatrix} EA_e & 0 \\ 0 & EA_\mu \end{pmatrix} \tag{23}$$

where:
$A_e \approx \left[\frac{G_F}{\sqrt{2}} V_{ud} F_\pi\right]^2 \left[\frac{m_e}{8m_\pi^2}\right] n_N$, $A_\mu \approx \left[\frac{G_F}{\sqrt{2}} V_{ud} F_\pi\right]^2 \left[\frac{m_\mu}{8m_\pi^2}\right] n_N$, $V_{ud}$ are the Cabibbo – Kobayashi- Maskawa (CKM) matrix elements $n_N = \int \frac{d^3q}{(2\pi)^3}$ is the nucleon number density and $q$ is the momentum transfer between neutrinos and quarks. $A_e \cong 3 \times 10^{-16}$  $A_e \cong 3 \times 10^{-14}$, see e.g.,[3]. The energy ε  is given by :

$$\epsilon = \frac{\Delta m_{12}^2}{2E} + E(A_\mu - A_e) + \frac{1}{8}\Delta B_{12}^2 E^3 a^2 + 4\hbar\Delta(\alpha E) \tag{24}$$

From Eqs. (16) and (24), the oscillation phase takes the following form :

$$\varphi = \varphi_{vac} + \varphi_{nuc} + \varphi_{LIV} \tag{25}$$

Again we find that similar to the vacuum, for very high energy neutrinos the contribution of  the mass-difference  term in Eq. (24), vanishes but contrary to the vacuum there is a non-vanishing contribution due to the interaction of neutrinos with quarks and  also a contribution from discreteness of space and VLI term.
In conclusion, the observation of  very high energy neutrinos oscillation  in vacuum may be a sign of  discreteness of space and VLI. But the oscillations of  very high energy neutrinos in normal and nuclear media are contributed from both interaction of neutrinos with matter and discreteness of  space. It is worth mentioning that we have assumed that the GUP parameter α , and the LIV free parameter  B, are energy dependent. If  they are constant, then their contributions to the osillation phase φ is zero in three cases discussed  in this work.

## References

[1]. A. F. Ali, S. Das, E. C. Vagenas,  Phys.Lett.B678 (2009) 497.
[2]. A. F. Ali, S. Das, E. C. Vagenas,  Phys. Lett. B 755 (2016) 17
[3]. I. Motie, She-Sheng Xue,  arXiv:1104.2837.
[4]. I. Motie, She-Sheng Xue, IJMPA 19 (2012) 1250104.

[5]. K. G. Wilson, Phys.Rev. D 10 (1974) 2445.

[6]. J. B. Kogut, Rev. Mod. Phys. 55 (1983) 775.
[7].C. Lunardini, A. Yu. Smirnov, Phys, Rev, D64,073006 (2001)